\documentstyle[aps,preprint]{revtex}
\begin{document}
\title{Proposal for an Experiment to Test a Theory of High
Temperature Superconductors}
\author{C. M. Varma}
\address{Bell Laboratories, Lucent Technologies \\
Murray Hill, NJ  07974}
\maketitle

\begin{abstract}
A theory for the phenomena observed in Copper-Oxide based 
high temperature
superconducting materials derives an elusive time-reversal 
and rotational
symmetry breaking order parameter for the observed pseudogap 
phase ending at a 
quantum-critical point near the composition for the highest 
$T_c$. An  experiment is proposed to observe such a symmetry 
breaking. It is shown that Angle-resolved Photoemission yields
 a current  density which is different for left and right
 circularly polarized photons.
The magnitude of the effect and its momentum dependence 
is estimated. Barring 
the presence of domains of the predicted phase an asymmetry 
of about 0.1 is predicted at low temperatures in moderately
 underdoped samples.

\end{abstract}

\newpage
\section{Introduction}
Despite twelve years of intesive experimental and theoretical studies
of copper-oxide based superconducting compounds, \cite{conferences} no
 consensus
about the fundamental physics or even about the minimum necessary
Hamiltonian to describe the phenomena has emerged.  One of the
 few theoretical ideas which has clearly survived experimental tests
is that at density $x \approx x_c$ near that for the maximum
superconducting transition temperature, the normal state is a
marginal Fermi-liquid (MFL)\cite{mfl}.  The MFL is characterized by a
scale-invariant particle-hole fluctuation spectrum which is only
very weakly momentum dependent.  One of the predictions of MFL
hypothesis is that the single-particle spectral function 
$G ( {\bf k} , \omega )$ has a nearly momentum-independent
self-energy proportional to $max ( | \omega |, T )$.  The
frequency and temperature dependence as well as the momentum
independence have been tested in angle-resolved photoemission
experiments.\cite{arpes}

The observed non-Fermi-liquid behavior near 
$x \approx x_c$ in resistivity, thermal conductivity, optical
conductivity, Raman scattering, tunneling spectra, and the Cu
nuclear relaxation rate follow from the MFL hypothesis.
The scale-invariance of the MFL fluctuations implies that a
quantum-critical-point (QCP) exists at $x = x_c$, near the
optimum composition.  One expects that, in two or 
three dimensions, the
QCP at $T = 0$ is the end-point of a phase of reduced symmetry as
 $x$ is varied. Similarly  a line of transitions or at least a
cross-over is expected at a finite temperature $T_p (x)$ terminating at $(x=x_c, T=0)$.  Indeed
the generic phase-diagram, Fig. (1), of the copper-oxide 
compounds around
$x \approx x_c$ displays such a topology.  Region I has MFL
properties dominated by quantum-fluctuations, Region III displays
properties characteristic of a Fermi-liquid, while Region III - the
pseudo-gap region displays a loss of low-energy excitations
compared to Region II.  Below the line $T_p (x)$ between regions I
and II, the single-particle spectrum displays lowered rotational
symmetry, while no translational
symmetry appears broken.  The superconductivity region sits
spanning the three-distinct normal state regions.

Fig. (1) may be compared to the topologically  similar
phase-diagram of some heavy-Fermion compounds, in which the line
$T_p (x)$ corresponds to an antiferromagnetic transition.\cite{mathur}  From
this point of view the crucial question in Cu-O compounds is the
symmetery in Region II of the so-called pseudo-gap phase.

A systematic theory\cite{varma1,varma2} starting with a general model for Cu-O
compounds provides an answer to this question.  The region (II)
in Fig. (1) is derived to be a phase in which a four-fold pattern
of current flows in the ground state in each unit cell as shown in
Fig. (2).  Time-reversal symmetry as well as rotational symmetry
is broken but the product of the two is conserved.  This phase has
been called the circulating-current (CC) phase.  Quantum
fluctuations about this phase are shown to have MFL fluctuations,
characteristic of Region I.  The same fluctuations promote ''$d$" or
generalized ''$s$"-state pairing depending on the Fermi-surface at
a given doping.

While a microscopic theory in agreement with most of the principal
experimental results has been presented, one can be confident of
the applicability of the theory only if the CC phase is directly
observed.  The CC phase has a very elusive-order parameter. The four-fold pattern of microscopic magnetic moments in
each unit cell changes the Bragg intensity for polarized neutrons
at certain pre-existing Bragg spots.  But the intensity for
nuclear-scattering at these Bragg spots is $O (10^4 )$ the
predicted magnetic intensity.  Muon spin-resonance ($\mu$-SR)
would be a possible probe, but the magnetic field from the current
pattern in Fig. (2) is zero at most symmetry points and along the
principal symmetry lines, where muons are known to sit
preferentially. Perhaps, an additional perturbation such as an external
magnetic field can be used to lower the symmetry at the sites preferred
by muons. In that case $\mu-SR$ could be used to search for the 
predicted phase. 

I propose here a new kind of experiment, which is a microscopic
analog of circular dichorism.  The idea is that ARPES at a
specific point near the Fermi-surface should have 
an electron yield which
is different for right circularly polarized and left circularly
polarized photons if the ground state has T-breaking of the form
shown in Fig. (2).  Further the relative intensity should
change in a
systematic fashion with the momentum around the Fermi-surface.

I present below the results of the calculation based on this idea
and then discuss the feasibility of the experiment. The idea itself is more general than the specific application to copper-oxides. Any time reversal breaking phase will in general yield a different current density for 
right and left circularly polarized photons. (But the characterestic 
signature of the state, revealed by the momentum dependence of the
asymmetry in the current for the left and right circular polarizations, 
must be calculated anew for each possibility.)
The experiment may for example  be tried to see if the superconducting      state of the compound $Sr_2RuO_4$ \cite{maeno} breaks time-reversal           symmetry.

\section{ARPES With Polarized Photons}

My object is to deduce the polarization and symmetry dependence of
ARPES current and a rough estimate of its magnitude.  For this
purpose, a simple calculation using tight-binding wave-functions
in the solid is sufficient.

Assume that a photon of energy $\omega$ shone on the crystal
produces a free-electron with momentum ${\bf p}$ and energy
$E_{\bf p}$ at the detector due to absorption of the photon by an
electronic state $| {\bf k} >$ inside the crystal of the crystal
momentum ${\bf k}$.  The momentum of the photon is assumed very
small compared to ${\bf k}$ and ${\bf p}$.  The current $J_{\bf p,\bf k}$      collected at the
detector for uniform illumination over a given area is \cite{ashcroft}
\begin{equation}
J_{\bf p,\bf k} = 2 \pi \, e \, f \left( \epsilon_{\bf k} \right) \left | 
\left< \: {\bf p} \, | H^{\prime} | \, {\bf k} \right> \:
\right|^2 \delta \left( E_{\bf p} - \epsilon_{\bf k} 
+ \hbar \omega \right)
\end{equation}
where $f ( \epsilon_{\bf k} )$ is the Fermi-function.

The primary contribution of the current is from the matrix element
\begin{equation}
\left< {\bf p} \, | H^{\prime} | \, {\bf k} \right> =
\frac{-ie}{2mc} \int d \; {\bf r} \: e^{i {\bf p} \cdot {\bf r}}
{\bf A} \cdot {\bf \nabla} \Psi_{\bf k} (r)
\end{equation}
where {\bf A} is the vector potential of the incident photons and 
$\Psi_{\bf k} ({\bf r})$ is the wave function of the state
$| {\bf k} >$.  There is a smaller contribution due to the gradient
of the potential at the surface which is briefly discussed at the
end.

\subsection{Wavefunctions}
The creation operator for the tight-binding wavefunctions for the conduction band of Cu-O
metals (assumed to be a two-dimensional metal) for the case that the
difference in energy of the
Cu-$d_{x^2 - y^2}$ level $\epsilon_d$ and the O-$p_{x,y}$ levels
$\epsilon_{\bf p}$ is much less than their hybridization energy,
and when the direct Oxygen-Oxygen hopping parameter 
$t_{pp}$ is set to zero are
\begin{equation}
| {\bf k}_o \rangle = 
\frac{d_k^+}{\sqrt{2}} \: + i\left(\: 
\frac{s_x \, p_{kx}^+ + s_y \, p_{ky}^+}{\sqrt{2} s_{xy}}\right)
\end{equation}
where $s_{x,y} = \sin k_{x,y} a/2$ and
$s_{xy}^2 = \frac{\sin^2 k_x a}{2} + \frac{\sin^2 k_y a}{2}$.
Spin labels have been suppressed.

$d_k^+$, $p_{kx,y}^+$ are respectively the creation operators for the basis wave-functions
\begin{eqnarray}
\phi_d({\bf k}) & = & \frac{1}{\sqrt{N}}  \sum_i e^{-i {\bf k \cdot R_i}} \: 
\phi_d ({\bf r - R_i}) , \nonumber \\
\phi_{p_{x,y}}({\bf k}) & = & \frac{1}{\sqrt{N}} \sum_i \:
e^{-i \bf k \cdot R_i} \: 
e^{-i k_{x,y}\frac{a}{2}} \phi_{p_{x,y}}({\bf r - R_i} - \frac{a_{x,y}}{2} )
\end{eqnarray}
where $\phi_d ({\bf r - R_i})$ is the $d_{x^2-y^2}$ atomic orbital at the
Cu-site $R_i$ and 
$\phi_{px} \left( {\bf r} - {\bf R}_i - \frac{a_x}{2} \right)$
is the $p_x$ wavefunction at the oxygen site at
${\bf R}_i + \frac{a_x}{2}$, etc.

In the circulating current phase, the conduction band
wave-function is modified to \cite{footnote}

\begin{equation}
|{\bf k} \rangle = \left( | {\bf k}_o \rangle +  \theta_0 
| {\bf k}_1 \rangle \right) / 
\sqrt{1 + \theta_o^2 \, s_x^2 \, s_y^2}
\end{equation}
where
\begin{equation}
| {\bf k}_1 \rangle \simeq s_x \, s_y
\left( s_y \: p_{kx}^+ - s_x \: p_{ky}^+ \right) / s_{xy}.
\end{equation}
In Eq. (5) $\theta_0$ characterises the strength of the              symmetry-breaking.

\subsection{Matrix-elements and Current}

  In order to evaluate the matrix element, Eq. (2), I write
\begin{equation}
\begin{array}{rl}
\phi_d ( {\bf r}) & = \psi_d (| {\bf r} | ) \:
\frac{(x^2 - y^2 )}{r^2} , \\
\phi_{p_\mu} ( {\bf r} ) & = \psi_p
( | {\bf r} |) \: \frac{\mu}{|r|}; \: \mu = x,y
\end{array}
\end{equation}
$\psi_d (| {\bf r}|)$ and $\psi_{p_\mu} (| r |)$ are characterized
by a fall-off distance {\it a} of the order of the atomic size.
Then
\begin{equation}
\begin{array}{rl}
{\bf \nabla} \phi_{p_\mu} ( {\bf r} ) & \approx
\left[ \frac{1}{{\it a}} \left( 
\frac{\hat{x}x + \hat{y}y + \hat{z}z}{| {\bf r} |} \right)
\frac{\mu}{| r |} +  \frac{\hat{\mu}}{| r |} \right]
\psi_p ( | r | ), \\
{\bf \nabla} \phi_d ({\bf r}) & \approx \left[ 
\frac{1}{{\it a}} \left( 
\frac{\hat{x}x + \hat{y}y + \hat{z}z}{| {\bf r}|} \right) 
\frac{( x^2 - y^2 )}{r^2} + 
\frac{2 (x \hat{x} - y \hat{y} )}{r^2} \right]
\psi_d ( | r | )  .
\end{array}
\end{equation}

We need the (two-dimensional) momentum distribution of the
wavefunctions in Eq. (2).  For this purpose define
\begin{equation}
\int d \, {\bf r} \: e^{i(p_x x + p_y y)}
\nabla_{\nu} \phi_d ( {\bf r}) \;  \equiv i \; 
f_d^{\nu} ( p_x , p_y ); \; \nu = x,y \: \:.
\end{equation}
Note that $f_d^x (p_x , p_y )$ can be written as the product of an
odd function of $p_x$ and an even function of $p_y$, etc.
Similarly,
\begin{equation}
\int d \, {\bf r} \: e^{i ( p_x x + p_y y )} \nabla_{\nu}
\phi_{p_\mu} (r) \; \equiv \; f_{p_\mu}^\nu (p_x , p_y )
\end{equation}
$f_{p_\mu}^{\mu} (p_x , p_y )$ is the product of an even function
of $p_x$ and an even function of $p_y$, whereas
$f_{p_\mu}^{\nu} (p_x , p_y )$ is the product of an odd function
of $p_x$ and an odd function of $p_y$. The definitions in Eqs.(9,10)
ensure that all the $f's$ are real. The $f(p)'s$ fall off for $p$
of order the inverse atomic size. Therefore for $p's$ in the first or
second Brillouin zone they are approximately constant.

In terms of these quantities, the matrix element in Eq. (2) is
calculated.  Consider the case of left and right circularly
polarized photons with vector potentials
${\bf A}_\ell$ and ${\bf A}_r$ respectively
\begin{equation}
{\bf A}_{\ell , r} \: = \: A 
\left( \hat{x} \pm i \, \hat{y} \right) \:.
\end{equation}
Then a straight forward calculation leads, to leading order 
in $\theta_0$ to
\begin{eqnarray}
 \langle {\bf p} \, | H^{\prime} | \, {\bf k} \rangle_{l,r} & =
(\frac{e}{2\sqrt{2}mc})A \sum_{Gx, Gy} \delta (\bf{ p - k -G} ) 
\left\{ \left( R_o ({\bf G}, {\bf p},{\bf k}) \pm iI_o 
({\bf G}, {\bf p},{\bf k}) \right) \right. \\
& \left. + \theta_0 \left(  \pm R_1 ({\bf G}, {\bf p},{\bf k})
+ i \, I_1 ({\bf G},{\bf p},{\bf k}) \right) \right\}
\end{eqnarray}
where ${\bf G} = (G_x , G_y )$ are the reciprocal vectors, and
\begin{equation}
R_o ({\bf G}, {\bf p},{\bf k}) = f_d^x ( {\bf p} )  +
 \left( g(G_y,k_x) 
f_{px}^y ( {\bf p}) + g(G_y,k_x) f_{py}^y ( {\bf p})\right) 
\end{equation}
\begin{equation}
I_o ( {\bf G}, {\bf p},{\bf k} ) = f_d^y ( {\bf p} ) - 
\left( g(G_x,k_y) f_{p_x}^x
( {\bf p} ) + g(G_y,k_x)
f_{p_y}^x ({\bf p}) \right)
\end{equation}
\begin{equation}
R_1 ({\bf G}, {\bf p},{\bf k} ) =
\left( \sin^2 \left( \frac{k_y a}{2}\right)
g(G_x,k_y) f_{p_x}^x ({\bf p}) -
\sin^2 \left( \frac{k_x a}{2}\right) g(G_y,k_x) f_{p_y}^x ({\bf p})\right)
\end{equation}
\begin{equation}
I_1 ({\bf G}, {\bf p},{\bf k},{\bf k}) = 
\left(sin^2 \left( \frac{k_y a}{2} \right) 
g(G_x,k_y) f_{p_x}^y 
({\bf p}) - \sin^2 \left(\frac{k_x a}{2} \right) g(G_y,k_x) f_{p_y}^y\right)
\end{equation}
In the above
\begin{equation}
g(r,s)= \frac{sin(ra/2)}{\sqrt{sin^2\left(ra/2\right)+sin^2\left(sa/2\right)}}
\end{equation}
In the usual experimental geometry, the contribution from each
{\bf G} is selected separately.  For a particular {\bf G}, the
current with polarization $\ell$ or $r$ to first order in
$\theta$ is 
\begin{equation}
J_{\ell , r} ({\bf G, p}) \, = \frac{e^2}{8m^2c^2}\, 
\left[ \left( R_o^2 + I_o^2 \right) \pm 2 \theta
\left( R_o \, R_1 \, + \, I_o I_1 \right) \right]
\end{equation}
so the relative asymmetry of the current,
\begin{equation}
\Xi ({\bf G}, {\bf p}) \equiv ( J_{\ell} - J_r ) / 
\frac{1}{2} ( J_{\ell} + J_r ) , \; \approx \; 
8 \, \theta_0 \; \left( R_o R_1 + I_o I_1 \right) /
\left( R_o^2 + I_o^2 \right)
\end{equation}

\section{Discussion of ARPES - Asymmetry}

Eqs. (19) and (20) are the principal result of the calculation.  It is
worthwhile noting several aspects of the predictions.  For
$G_x = G_y = 0$, when only the d-orbitals contribute to the
photo-current, $\Xi = 0$ for all ${\bf k}$. For
$G_x = G_y$, the asymmetry vanishes along the zone-diagonal
$k_x = k_y$ and is maximum for the zone-boundaries
$( k_x a = \pi , k_y a = 0 )$; $(k_x a = 0, k_y a = \pi )$ with a 
smooth variation in between. Asymmetry patterns for other $G's$ may be 
obtained from Eqs. (14)-(17).

In Ref. (6), $\theta_0$ is estimated to be 
$O (10^{-1} ) (x_c - x)^{1/2}$ for $x \leq x_c$.  So at
$x_c - x \approx 5 \times 10^{-2}$, the asymmetry is predicted to
be $O (10^{-1})$, at $T \approx T_p (x)$.

The proposed experiment is to measure the ARPES current in under-doped
samples for a
fixed relative geometry of the incident photon-beam, crystalline
surface, and the detector to select a {\bf p} and {\bf G} and then
simply switch the polarization of the incident photons, and
measure the current again.  The experiment should then be repeated
for different {\bf p} and {\bf G}. The effect should set in for
 temperatures $T\lesssim T_p(x)$ and have a momentum dependence predicted
by Eq. (20) and Eqs. (14)-(17).

The principal difficulty of the experiment is the possible
presence of domains of the CC phase.  The domains consist of
regions in which $( \theta_0 )$ in the wave-function (5) is
replaced by $( -\theta_0 )$.  This leads to a mutual switching of
the pattern of the current within the unit cells (and a current
flow along the domain boundary).  The effect calculated in Eq.
(20) to linear order in $\theta_0$ then averages to zero for equal
number of the two-kinds of domains in the surface area $S$
from which the current is collected.  If the characteristic domain           size is $D$, an
effect proportional to $(D/S)^{1/2} \; \theta_0$ is still to be
expected.  Also, Eqn. (1) yields asymmetry terms proportional to
$\theta_0^2$, which are not affected by the domains.  However,
these may be too small to be observable.

In the above, circularly polarized photons, with the plane of
polarization along the surface of the crystal have been
considered.  There is also an effect linear in $\theta_0$
for photons linearly polarized along the normal to the surface due to
the potential gradient at the surface $(\nabla V)_s$.  This effect,
proportional to $(\nabla V)^2_s$ changes sign for a given
$G_x = G_y$, as $p_x$ and $p_y$ are interachanged in a d-wave like
fashion. Observation of this effect requires rotating the sample. It is also
affected by domains.

\newpage
\section*{Figure Captions}
\begin{itemize}
\item[Fig. (1):]
Generic phase-diagram of the cuprates for hole-doping.  
\item[Fig. (2):]
Current pattern predicted in phase II of Fig. (1) in References (5) and (6).
\end{itemize}

\begin{references}
\bibitem{footnote}
 I have picked a different gauge for the wave-functions compared to Ref.(6) , but one  which is the most convenient for calculation reported here. Also, compared to Ref.(6), a coefficient has been absorbed 
on the definition of $\theta_0$. In Ref.(6) additional modifications to the 
wavefunctions are discussed, which are not considered here. They produce
further corrections.
\end{references}
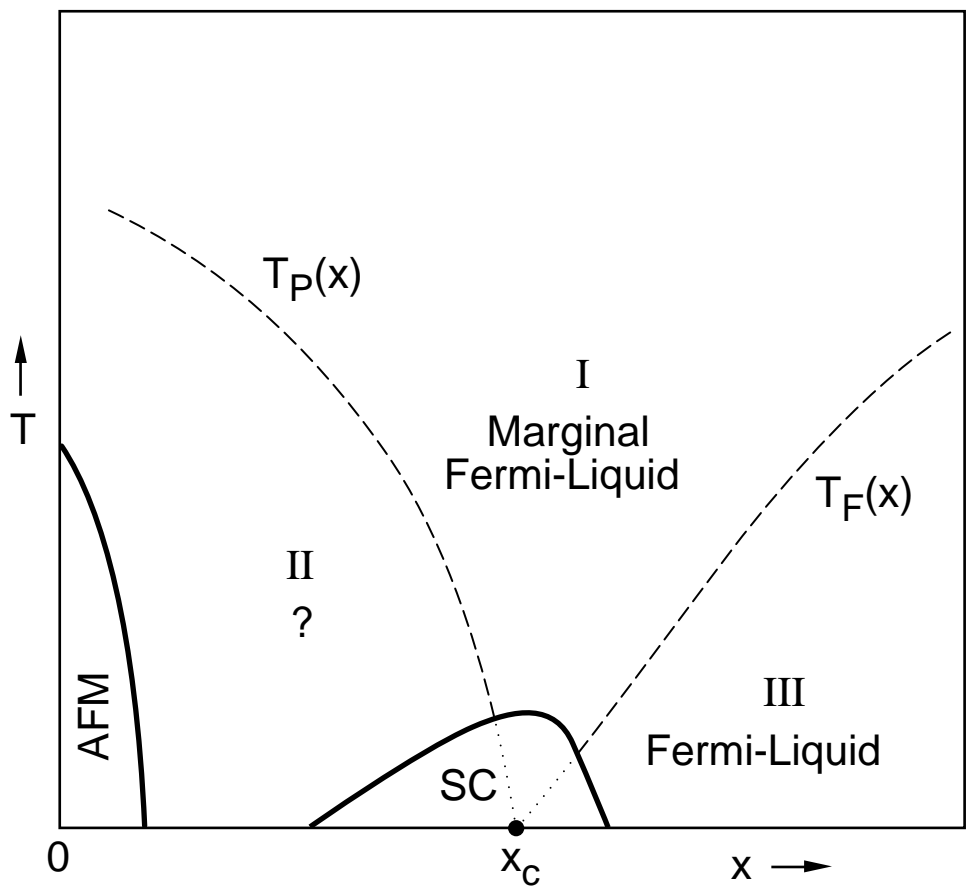
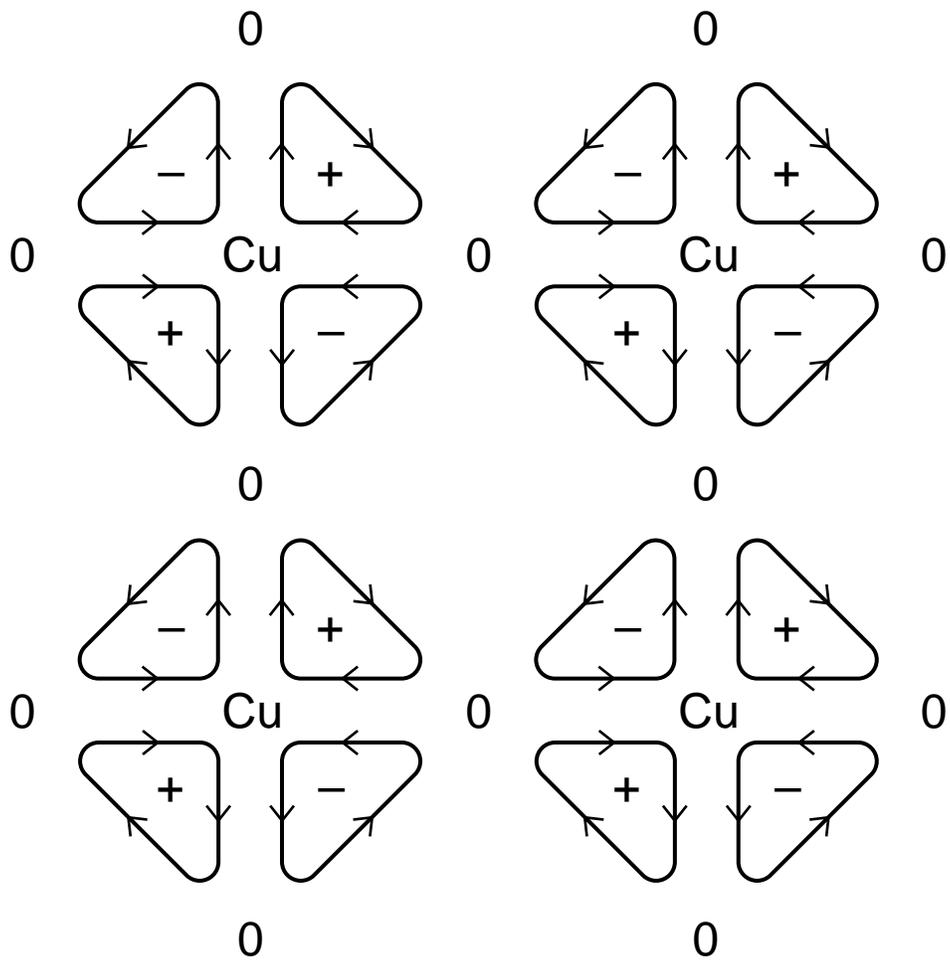
\end{document}